\documentclass[conference]{IEEEtran}
\IEEEoverridecommandlockouts
\usepackage{cite}
\usepackage{amsmath,amssymb,amsfonts}
\usepackage{algorithmic}
\usepackage{graphicx}
\usepackage{textcomp}
\usepackage{xcolor}
\usepackage{hyperref}
\hypersetup{
    colorlinks=true,
    linkcolor=red,
    urlcolor=blue,
    citecolor=black,
}

\def\BibTeX{{\rm B\kern-.05em{\sc i\kern-.025em b}\kern-.08em
    T\kern-.1667em\lower.7ex\hbox{E}\kern-.125emX}}
\begin{document}

\title{Preference vs. Performance: EEG-Based Classification of Learner Engagement in Multimodal Instruction}

\author{
\IEEEauthorblockN{
Sri Jahnavi Adusumilli$^{1}$,
Deepak Giri$^{1,2}$,
Pallavi Vaswani$^{1}$,
Pallavi Singh$^{1}$,\\
Megha Moncy$^{1}$,
Lalitha Pranathi Pulavarthy$^{1}$,
Saptarshi Purkayastha$^{1}$
}
\IEEEauthorblockA{$^{1}$Department of Biomedical Engineering and Informatics,
Indiana University, Indianapolis, IN 46202 USA
}
\IEEEauthorblockA{$^{2}$Department of Media and Information,
Michigan State University, East Lansing, MI 48918 USA
}
}

\maketitle

\begin{abstract}
Effective adaptive instructional systems require robust measures of learner engagement that go beyond static user profiles. This study employs Multimodal Learning Analytics (MMLA) to investigate the relationship between self-reported instructional modality preferences and objective neurophysiological markers of engagement. Thirty-seven participants engaged with learning content delivered via varying modalities (visual, auditory, reading/writing, kinesthetic). We captured real-time neural activity using two EEG devices: the Emotiv EpocX (14 channels, 128 Hz) and OpenBCI (16 channels, 125 Hz). Preferences were assessed using the VARK questionnaire.
Consistent with literature challenging the ``meshing hypothesis,'' aligning instructional modality with stated preferences did not significantly predict performance gains. However, spectral analysis of EEG data revealed divergent engagement patterns: when content aligned with preferences, distinct neural activity patterns emerged in theta and alpha frequency bands-markers associated with attention and cognitive processing. These signals were used to train a binary logistic regression classifier, achieving a mean accuracy of 83.21\% with OpenBCI and 56.27\% with Emotiv EpocX. These findings suggest that while self-reported preferences may not dictate learning outcomes, they significantly influence neurophysiological engagement, offering a viable, non-invasive input for adaptive educational algorithms.
\end{abstract}


\section{Introduction}
Understanding how learners engage with different instructional formats represents a persistent challenge in educational research. While the concept of ``learning styles'' - the notion that individuals have preferred modalities through which they best acquire information - has been influential in educational practice, substantial empirical evidence has accumulated questioning whether matching instruction to stated preferences actually improves learning outcomes \cite{pashler2008learning}. Specifically, the ``meshing hypothesis'' - that aligning instruction with a learner's preferred modality optimizes performance - has failed to find consistent support in controlled experiments \cite{willingham2015scientific,rogowsky2020providing}, including studies that used the VARK inventory itself \cite{husmann2019another}. Nevertheless, the question of how instructional modalities affect learner engagement, attention, and motivation remains educationally significant and distinct from claims about learning optimization. Critically, this study distinguishes between two separate claims often conflated in prior research: (1) that individuals differ in their preferences for how information is presented, and (2) that matching instruction to these preferences improves academic performance. Our investigation focuses primarily on the former, examining whether neurophysiological measures can detect differential engagement responses to instructional modalities, even if those differences do not translate to performance gains.

Current instructional design often adopts a one-size-fits-all approach that may not optimize engagement for all learners. While evidence does not support the claim that matching instruction to learning styles improves academic performance, research does suggest that learner engagement - a multidimensional construct encompassing behavioral, emotional, and cognitive components \cite{fredricks2004school}, is a robust predictor of academic outcome \cite{lei2018relationships}. Engaged learners demonstrate sustained attention, positive affect toward learning activities, and deeper cognitive processing \cite{d2012dynamics}. Understanding how instructional modality influences these engagement dimensions could inform more effective adaptive learning systems, wherein the system adapts to maintain user attention rather than to validate a specific learning style theory.

Engagement in learning contexts manifests through both subjective experience and objective physiological markers. Prior research has employed diverse methodological approaches to capture this construct. Self-report measures remain the most common approach, yet they are limited by retrospective bias and social desirability effects \cite{azevedo2015defining,dmello2017advanced}. Behavioral measures, such as time-on-task and interaction patterns logged by learning systems \cite{purkayastha2019critical}, provide implicit indices but often fail to capture internal cognitive states \cite{henrie2015measuring}. To address these limitations, physiological measures including electroencephalogram (EEG), heart rate variability, and electrodermal activity have been employed to assess engagement continuously and unobtrusively \cite{giannakos2019multimodal} \cite{darnell2019student}. EEG, in particular, provides a non-invasive method to measure brain electrical activity, offering objective insights into attentional allocation and cognitive workload that complement self-report data \cite{klimesch1999eeg} \cite{berka2007eeg}, and low-cost headsets have made such recordings feasible during computer-based instruction and in live classrooms \cite{wang2014exploratory,dikker2017brain}.

Specific EEG frequency bands have been associated with distinct cognitive and attentional states relevant to learning engagement. Theta oscillations (4–8 Hz), particularly over frontal regions, correlate with sustained attention, cognitive control, and memory encoding \cite{klimesch1999eeg,cavanagh2014frontal}. Alpha rhythms (8–13 Hz) demonstrate event-related desynchronization during active cognitive processing and are inversely related to cortical activation \cite{pfurtscheller1999event}. Pope et al. proposed the ratio $\beta/(\alpha+\theta)$ as an index of task engagement \cite{pope1995biocybernetic}, whereas frontal theta increases and parietal alpha decreases with working-memory load \cite{gevins1997high}, so theta/alpha measures are better read as indices of workload than of engagement. Beta oscillations (13–30 Hz) are associated with active concentration and alertness \cite{ray1985eeg}. Prior work using EEG to classify engagement states has achieved accuracies ranging from 70\% to 90\% depending on task, population, and EEG system \cite{ko2017sustained} \cite{muhl2014eeg}. A related study classified learners along the active--reflective dimension of the Felder--Silverman learning-style model from 14-channel Emotiv EEG, reaching 71.2\% accuracy with 14 participants \cite{zhang2021design}; however, it did not examine the Visual, Aural, Read/Write, Kinesthetic (VARK) modalities, and the finding awaits replication with larger samples and clearer theoretical framing around engagement rather than learning style validation.

Despite growing interest in neurophysiological measurement in educational contexts, several gaps remain. First, few studies have examined whether EEG can detect differential engagement when instructional modality matches or mismatches learner preferences-and critically, whether such neural differences predict learning outcomes. Second, the relationship between subjective preference and objective engagement remains underexplored; individuals may prefer modalities that do not actually optimize their neurophysiological engagement. Third, consumer-grade EEG devices vary substantially in signal quality \cite{sawangjai2020consumer}, and comparative validation across devices in educational contexts is needed. The present study addresses these gaps by: (1) collecting both self-reported preferences (using Fleming's VARK questionnaire \cite{fleming1992not}, whose psychometric structure has itself been questioned \cite{leite2010attempted}) and objective EEG data, (2) examining both engagement markers and learning outcomes as separate dependent variables, and (3) comparing classification performance across research-grade Open-Source Brain-Computer Interface (OpenBCI) and consumer-grade (Emotiv EpocX) EEG systems. We hypothesize that EEG spectral features can distinguish matched from mismatched instructional conditions, while remaining agnostic about whether such matching could improve outcomes.

\section{Methodology}
\subsection{Study Design}
This study utilizes a quasi-experimental Multimodal Learning Analytics (MMLA) design to investigate the intersection of learner preference and neurophysiological engagement. The research adopts a mixed-methods approach, triangulating self-reported psychometric data with objective real-time EEG measurements. The core methodological contributions include: (1) the ecological collection of neurophysiological data within an authentic graduate-level classroom environment, (2) a comparative validation of consumer-grade (Emotiv EpocX) versus research-grade (OpenBCI) EEG systems for educational monitoring, and (3) the pilot implementation of a closed-loop engagement detection pipeline.

\subsection{Participants}
\begin{figure}[t]
    \centering
    \includegraphics[width=0.4\textwidth]{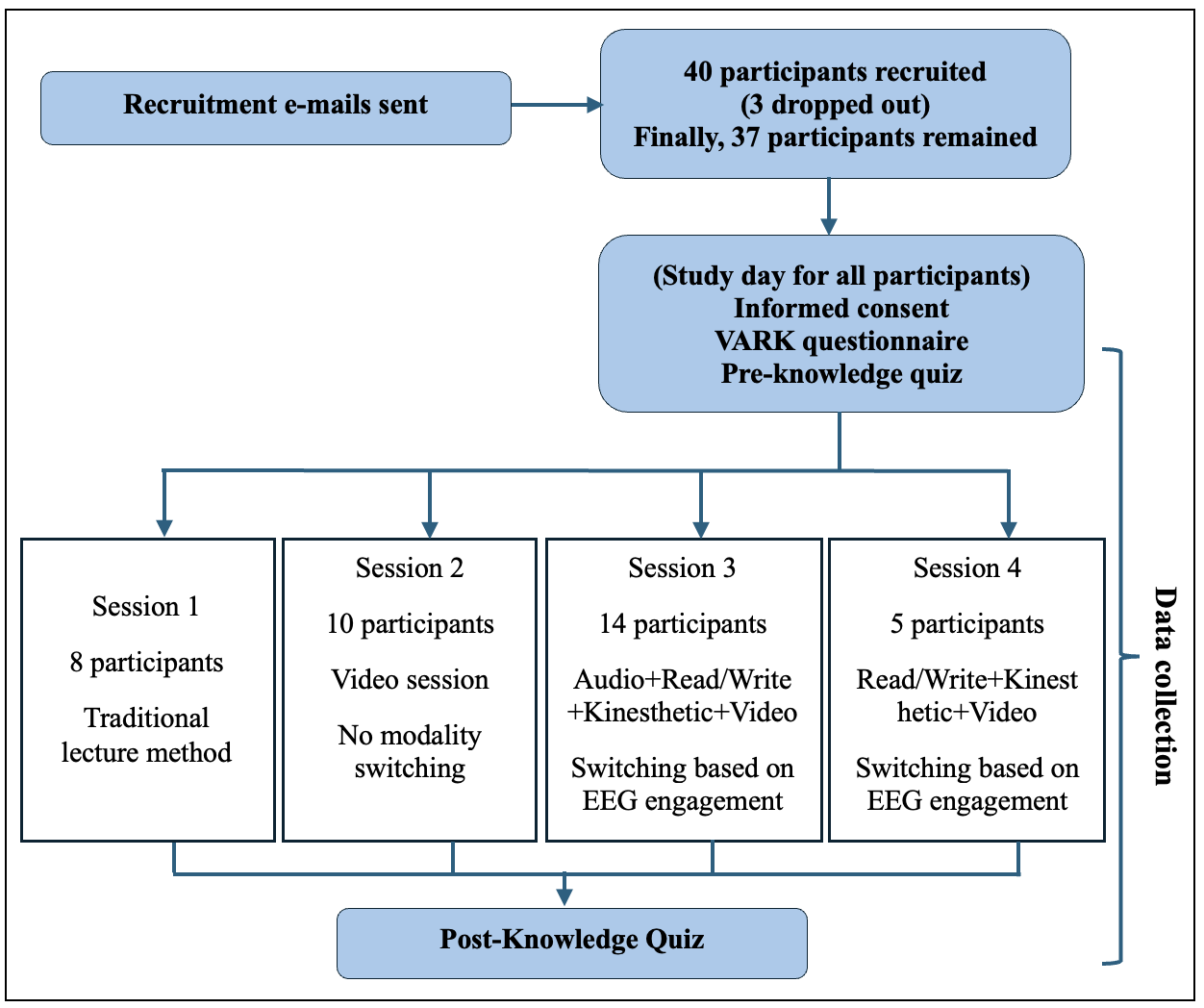}
    \caption{Schematic of the study design and participant distribution across sessions.}
    \label{fig:methodology}
\end{figure}

Following Institutional Review Board approval (\#17732), a total of 37 graduate students in Informatics were recruited from Indiana University Indianapolis. To accommodate the constraints of high-density EEG collection in a live classroom, participants were distributed across four distinct data collection sessions ($n_1 = 8$, $n_2 = 10$, $n_3 = 14$, $n_4 = 5$), as illustrated in Fig.~\ref{fig:methodology}. 

Within this cohort, a subset of eight participants ($n_{EEG}=8$) was selected for neurophysiological monitoring. In each session, two participants were instrumented: one wearing the OpenBCI system and one wearing the Emotiv EpocX. This sampling strategy was designed to assess device feasibility while minimizing classroom disruption.

\subsection{Experimental Procedure}
The experimental protocol consisted of three phases:
\begin{enumerate}
    \item \textbf{Baseline Assessment:} Participants completed the VARK questionnaire (Version 8.01) to establish their dominant learning modality preference (Visual, Aural, Read/Write, or Kinesthetic). A pre-assessment of Natural Language Processing (NLP) domain knowledge (10 items, 0--100 scale) was administered to establish a performance baseline.
    \item \textbf{Instructional Intervention:} Each session consisted of a 30-minute instructional module on NLP. Session 1 utilized a traditional lecture format (Control). Sessions 2--4 employed an adaptive MMLA-informed instruction model, where the instructional delivery modality was dynamically altered in real-time when aggregate EEG engagement indices fell below a pre-set threshold.
    \item \textbf{Post-Assessment:} Immediately following instruction, participants completed a post-assessment to measure learning gains and retention.
\end{enumerate}

\subsection{EEG Acquisition and Preprocessing}
\begin{figure}[!btp]
    \centering
    \includegraphics[width=0.4\textwidth]{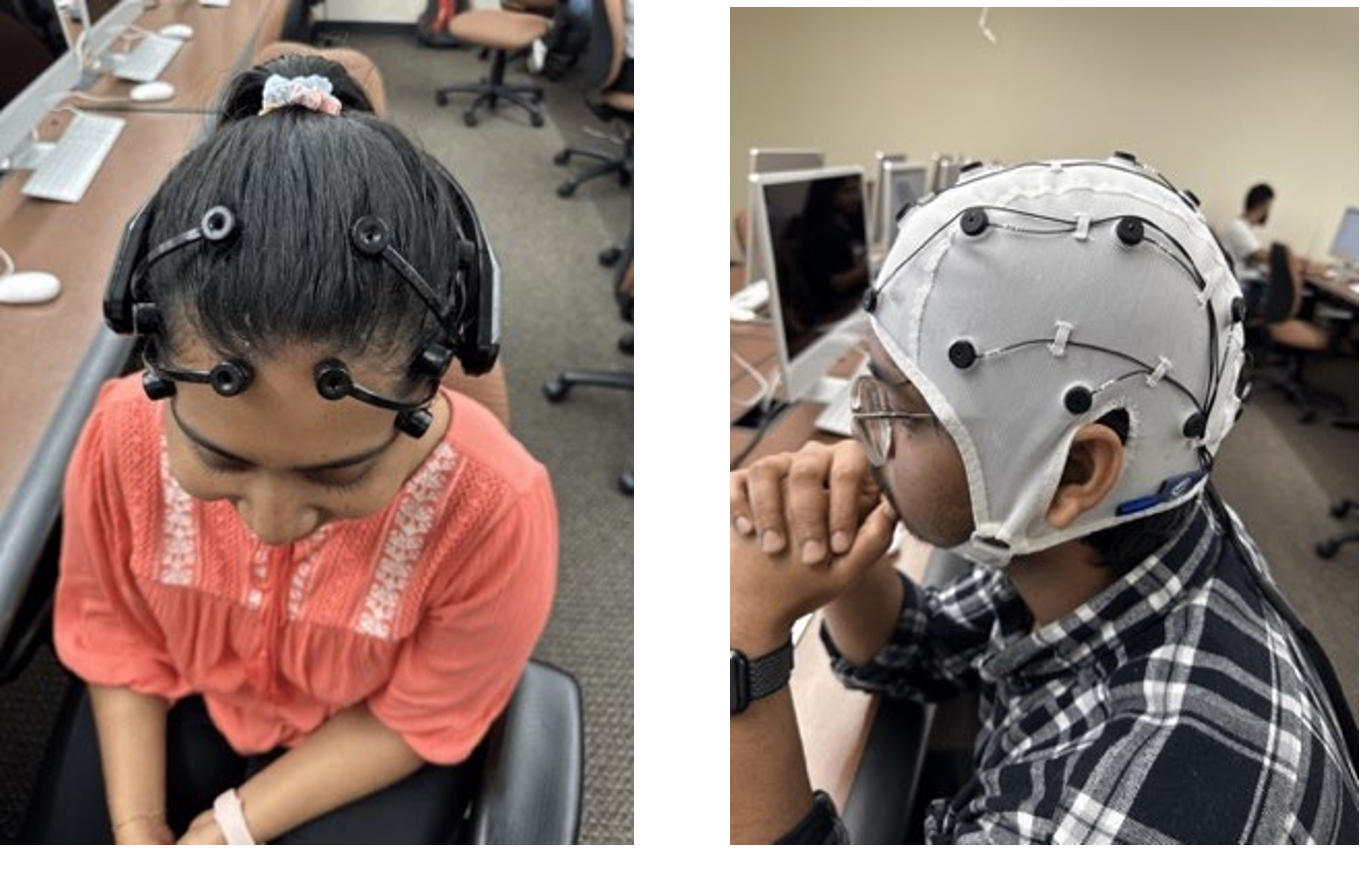}
    \caption{Experimental setup showing usage of Emotiv (left) and OpenBCI (right) EEG devices.}
    \label{fig:devices}
\end{figure}

Neurophysiological data were captured using two distinct systems to allow for cross-device comparison:
\begin{itemize}
    \item \textbf{Emotiv EpocX:} A consumer-grade headset with 14 saline-soaked felt sensors (128 Hz sampling rate), positioned according to the international 10--20 system (AF3, F7, F3, FC5, T7, P7, O1, O2, P8, T8, FC6, F4, F8, AF4).
    \item \textbf{OpenBCI Cyton+Daisy:} A research-grade system with 16 wet active electrodes (125 Hz sampling rate), providing higher signal fidelity (see Fig.~\ref{fig:devices}).
\end{itemize}
Data processing was performed using EEGLAB (v2025.0) \cite{delorme2004eeglab} within the MATLAB environment. Continuous EEG data were bandpass filtered (0.5--50 Hz) to remove DC drift and high-frequency noise. A 60 Hz notch filter was applied to suppress power line interference. Artifacts were removed via visual inspection and Independent Component Analysis (ICA) to isolate ocular and muscular artifacts. The cleaned data were segmented into 1-second non-overlapping epochs. Fast Fourier Transform (FFT) was applied to extract spectral power densities in five canonical frequency bands: delta (1--4 Hz), theta (4--8 Hz), alpha (8--13 Hz), beta (13--30 Hz), and gamma (30--45 Hz). Principal Component Analysis (PCA) was subsequently applied to the feature vectors to reduce dimensionality and mitigate the risk of overfitting.

\subsection{Classification Framework}
To validate the utility of EEG for engagement detection, a binary classification task was defined. The objective was to distinguish between epochs where the instructional modality \textit{matched} the participant's dominant VARK preference versus epochs where it \textit{mismatched}. This operationalizes the research question of whether preference alignment manifests in detectable neurophysiological signatures.

Binary Logistic Regression was selected as the classifier due to its interpretability and suitability for smaller datasets. To rigorously assess the model's ability to generalize to new users (a critical requirement for BCI applications, since sensor-based models that perform well within one group of people can transfer poorly to another \cite{aswar2023generalizability}), Leave-One-Participant-Out Cross-Validation (LOPOCV) was employed. In each fold, all epochs from one participant served as the test set, while the remaining participants formed the training set. Within training folds, the minority class was oversampled to ensure class balance. Model performance was evaluated using Accuracy, Precision, Recall, and F1-score. Bootstrap resampling (1,000 iterations) was applied to generate 95\% confidence intervals.

\subsection{Statistical Analysis of Learning Outcomes}
Behavioral data analysis focused on the relationship between modality alignment and learning gains. Normality was assessed using the Shapiro--Wilk test. Due to violations of normality assumptions ($p < .05$), non-parametric tests were utilized. A Wilcoxon signed-rank test evaluated overall pre- to post-session knowledge gains. Mann–Whitney U tests compared learning gains between ``matched'' and ``mismatched'' conditions. Effect sizes were calculated using rank-biserial correlation ($r_{rb}$). All analyses were conducted with $\alpha = .05$, identifying results as exploratory given the limited sample size.

\subsection{Methodological Limitations}
Several limitations constrain the generalizability of these findings. First, the neurophysiological subsample ($n = 8$) limits statistical power; results should be viewed as a proof-of-concept for the MMLA pipeline rather than definitive population estimates. Second, the between-session design introduces potential confounds regarding time-of-day effects and session order. Third, the lack of continuous self-reported engagement (e.g., via experience sampling) precludes the cross-validation of EEG indices against subjective momentary experience. These constraints are further addressed in the Discussion section.

\section{Results}
\subsection{Modality Preference Distribution}
\begin{figure}[t]
    \centering
    \includegraphics[width=0.35\textwidth]{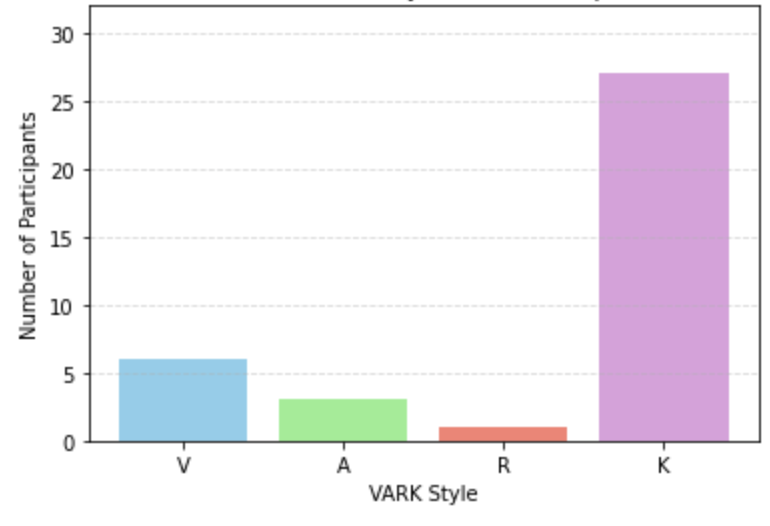}
    \caption{Distribution of dominant VARK modality preferences among participants ($N=37$).}
    \label{fig:learning_styles}
\end{figure}

The VARK assessment identified Kinesthetic as the predominant learning modality preference within the sample ($n = 27, 72.97\%$), followed by Visual ($16.22\%$), Auditory ($8.11\%$), and Reading/Writing ($2.70\%$). A significant subset of participants exhibited multimodal profiles, indicating secondary preferences alongside their dominant mode (see Fig.~\ref{fig:learning_styles}). This distribution reflects authentic learner variability, though the skew toward Kinesthetic preferences necessitates cautious interpretation of modality-specific subgroups.

\subsection{Behavioral Learning Outcomes}
Consistent with literature challenging the ``meshing hypothesis,'' alignment between instructional modality and learner preference did not reliably predict short-term learning gains. Post-session assessment scores indicated high variability: $37.84\%$ of participants improved, $21.62\%$ maintained their score, and $40.54\%$ showed a decline. A Wilcoxon signed-rank test confirmed no statistically significant overall pre-to-post difference ($W = 189.5, p = 0.536$).

Exploratory subgroup analyses using Mann–Whitney U tests suggested potential modality-specific effects. Visual learners in visual-dominant sessions ($U = 141.0, p = 0.023$) and Kinesthetic learners in kinesthetic-dominant sessions ($U = 210.0, p = 0.031$) showed marginal performance improvements. However, these uncorrected p-values should be interpreted as trend-level signals rather than confirmatory evidence, particularly given the small subgroup sizes.

\subsection{Neurophysiological Classification}
In contrast to the behavioral null results, neurophysiological signal analysis revealed distinct patterns of engagement. The binary logistic regression classifier successfully distinguished between preference-matched and mismatched instructional epochs, particularly when using research-grade hardware.

As detailed in Table~\ref{tab:classification}, the OpenBCI system achieved a mean classification accuracy of \textbf{83.21\%} (95\% CI: [82.42, 84.04]), substantially exceeding the theoretical chance baseline of 50\%. This indicates that spectral power features, specifically in theta ($4-8$ Hz) and alpha ($8-13$ Hz) bands, contain discriminative information regarding the learner's alignment with the instructional mode.

\begin{table}[htbp]
\renewcommand{\arraystretch}{1.2}
\centering
\caption{Engagement Classification Performance by EEG Device}
\label{tab:classification}
\begin{tabular}{|l|c|c|c|c|}
\hline
\textbf{Device} & \textbf{Accuracy} & \textbf{95\% CI} & \multicolumn{2}{c|}{\textbf{Recall}} \\
\cline{4-5}
 & (\%) & (\%) & Matched & Mismatched \\
\hline
OpenBCI & 83.21 & 82.42--84.04 & 81.86\% & 84.22\% \\
\hline
Emotiv EpocX & 56.27 & 55.10--57.52 & --- & --- \\
\hline
\end{tabular}
\end{table}

\subsection{Device Comparison and Signal Fidelity}
A marked performance gap was observed between hardware systems. While the OpenBCI system supported robust classification, the consumer-grade Emotiv EpocX yielded only marginal predictive power (56.27\%), hovering just above chance. This discrepancy likely reflects the superior signal-to-noise ratio of wet-electrode systems (OpenBCI) compared to saline-sensor arrays (Emotiv) in uncontrolled classroom environments \cite{sawangjai2020consumer}, suggesting that low-cost consumer EEG may currently lack the fidelity required for nuanced affective state detection without extensive artifact correction.

\subsection{Summary of Principal Findings}
Two diverging trends emerge from the data:
\begin{enumerate}
    \item \textbf{Behavioral Dissociation:} Alignment of instruction to self-reported preference did not yield significant aggregate performance gains, reinforcing skepticism regarding the pedagogical utility of learning styles for outcome optimization.
    \item \textbf{Neurophysiological Sensitivity:} Despite the lack of performance impact, preference alignment induced detectable shifts in neural oscillation patterns. The high classification accuracy ($>83\%$) suggests that ``preference'' manifests as a distinct physiological engagement state, even if that state does not causally drive immediate test score improvement.
\end{enumerate}

\section{Discussion}
\subsection{The Dissociation of Engagement and Performance}
The primary contribution of this study is the empirical dissociation between neurophysiological engagement and academic performance. Consistent with the ``meshing hypothesis'' critiques by others \cite{pashler2008learning,rogowsky2020providing}, our results found that aligning instruction to stated preferences did not yield significant learning gains. However, the neurophysiological data tell a different story: the classifier achieved 83.21\% accuracy in distinguishing matched from mismatched conditions. 

This suggests a critical nuance: subjective preference heavily influences neural engagement states, even if it does not optimize information encoding. This parallels behavioral evidence that learners' style preferences shape their judgements of learning without improving what they actually recall \cite{knoll2017learning}. For MMLA researchers, this implies that ``learning styles'' may be invalid as a pedagogical taxonomy but highly valid as a feature for affective computing. Adaptive systems might utilize preference-based content not to improve test scores directly, but to ``warm up'' the learner's neural state-maintaining attention and motivation (theta/alpha regulation) so that difficult cognitive work can occur.

\subsection{Device Feasibility in Educational Settings}
The substantial performance gap between the research-grade OpenBCI system (83.21\% accuracy) and the consumer-grade Emotiv EpocX (56.27\%) offers a sobering practical insight for EdTech scaling. While consumer saline-electrode headsets offer ease of use, our findings suggest they struggle with the signal-to-noise ratio inherent in ambulatory classroom settings. Movement during kinesthetic activities compounds this problem, and it is precisely the setting in which wearable inertial sensors can complement EEG \cite{yin2026multimodal}.

The reliance on saline-soaked felt sensors (Emotiv) likely resulted in higher impedance fluctuations compared to the wet electrode gel interface (OpenBCI), particularly over the parietal regions critical for alpha rhythm detection. Future implementations may prioritize semi-dry or active-electrode systems, bridging the gap between fidelity and usability.

\subsection{Contextualizing Preferences}
The skew toward Kinesthetic preferences (72.97\%) likely reflects the disciplinary context of the sample (Applied Informatics) rather than a general population trait; health informatics programs at this institution emphasize hands-on work in live clinical information systems \cite{sunchu2024lessons,purkayastha2019critical}. This reading is consistent with evidence that engagement is malleable and responsive to context \cite{fredricks2004school}, and that students' reported VARK preferences often do not match the study strategies they actually use \cite{husmann2019another}, suggesting that ``learning styles'' are likely context-dependent habits rather than fixed neurological traits. The high classification success, despite this skew, suggests that the EEG markers of engagement are robust, even within a homogeneous population.

\section{Limitations}
Findings should be interpreted within three key constraints. First, the neurophysiological subsample ($n=8$) limits statistical power, positioning these results as a proof-of-concept for the MMLA pipeline rather than a population-level generalization; with so few participants, the theoretical 50\% chance level is also an unreliable benchmark, and permutation testing would be needed to confirm above-chance decoding \cite{combrisson2015exceeding}. Second, the sample was heavily skewed toward Kinesthetic preferences (72.97\%), reflecting the specific disciplinary context of Informatics, which may limit generalizability to other fields. Third, the study utilized unimodal physiological sensing; the absence of concurrent eye-tracking or electrodermal activity data precludes the ``multimodal fusion'' validation that is becoming standard in engagement research \cite{giannakos2019multimodal}. Fusion benefits can be task-dependent: in a 16-channel EEG and inertial measurement unit (IMU) study of ten motor activities, IMU outperformed EEG on most tasks while EEG was stronger on rhythmic movement, and logistic-regression late fusion raised worst-task accuracy from roughly 87\% to 96.8\% \cite{yin2026multimodal}. Kinesthetic instruction, which combines cognitive load with physical movement, is a natural candidate for similar EEG--IMU fusion.

\section{Future Work}
Future research should move from detection to \textit{adaptation}. We propose developing closed-loop systems that dynamically alter instructional modality in real-time when EEG engagement indices drop, rather than relying on static pre-profiles, building on closed-loop EEG biofeedback systems for learners \cite{kosmyna2019attentivu}. Longitudinal designs are also required to determine if the engagement benefits of preference alignment persist after the novelty of the technology fades.

\section{Conclusion}
This study demonstrates that while self-reported preferences (VARK) do not dictate learning outcomes, they significantly modulate neurophysiological engagement. We successfully trained a binary classifier to distinguish preference-aligned instruction with $>$83\% accuracy, validating the presence of distinct neural engagement states. These findings support a paradigm shift in adaptive learning: rather than categorizing students into static ``styles'' to improve grades, educational technologies should treat preferences as inputs for engagement optimization, utilizing real-time neural feedback to maintain learner attention regardless of the pedagogical strategy.

\bibliography{references}

@article{pashler2008learning,
  title={Learning styles: Concepts and evidence},
  author={Pashler, Harold and McDaniel, Mark and Rohrer, Doug and Bjork, Robert},
  journal={Psychological Science in the Public Interest},
  volume={9},
  number={3},
  pages={105--119},
  year={2008},
  publisher={SAGE Publications},
  doi={10.1111/j.1539-6053.2009.01038.x}
}

@article{rogowsky2020providing,
  title={Providing instruction based on students' learning style preferences does not improve learning},
  author={Rogowsky, Beth A. and Calhoun, Barbara M. and Tallal, Paula},
  journal={Frontiers in Psychology},
  volume={11},
  pages={164},
  year={2020},
  publisher={Frontiers Media SA},
  doi={10.3389/fpsyg.2020.00164}
}

@article{fleming1992not,
  title={Not another inventory, rather a catalyst for reflection},
  author={Fleming, Neil D. and Mills, Colleen},
  journal={To Improve the Academy},
  volume={11},
  number={1},
  pages={137--155},
  year={1992},
  doi={10.1002/j.2334-4822.1992.tb00213.x}
}

@article{willingham2015scientific,
  title={The scientific status of learning styles theories},
  author={Willingham, Daniel T. and Hughes, Elizabeth M. and Dobolyi, David G.},
  journal={Teaching of Psychology},
  volume={42},
  number={3},
  pages={266--271},
  year={2015},
  publisher={SAGE Publications},
  doi={10.1177/0098628315589505}
}

@article{fredricks2004school,
  title={School engagement: Potential of the concept, state of the evidence},
  author={Fredricks, Jennifer A. and Blumenfeld, Phyllis C. and Paris, Alison H.},
  journal={Review of Educational Research},
  volume={74},
  number={1},
  pages={59--109},
  year={2004},
  publisher={SAGE Publications},
  doi={10.3102/00346543074001059}
}

@article{lei2018relationships,
  title={Relationships between student engagement and academic achievement: A meta-analysis},
  author={Lei, Hao and Cui, Yunhuo and Zhou, Wenye},
  journal={Social Behavior and Personality: An International Journal},
  volume={46},
  number={3},
  pages={517--528},
  year={2018},
  publisher={Scientific Journal Publishers},
  doi={10.2224/sbp.7054}
}

@article{d2012dynamics,
  title={Dynamics of affective states during complex learning},
  author={{D'Mello}, Sidney and Graesser, Art},
  journal={Learning and Instruction},
  volume={22},
  number={2},
  pages={145--157},
  year={2012},
  publisher={Elsevier},
  doi={10.1016/j.learninstruc.2011.10.001}
}

@article{azevedo2015defining,
  title={Defining and measuring engagement and learning in science: Conceptual, theoretical, methodological, and analytical issues},
  author={Azevedo, Roger},
  journal={Educational Psychologist},
  volume={50},
  number={1},
  pages={84--94},
  year={2015},
  publisher={Taylor \& Francis},
  doi={10.1080/00461520.2015.1004069}
}

@article{klimesch1999eeg,
  title={{EEG} alpha and theta oscillations reflect cognitive and memory performance: A review and analysis},
  author={Klimesch, Wolfgang},
  journal={Brain Research Reviews},
  volume={29},
  number={2--3},
  pages={169--195},
  year={1999},
  publisher={Elsevier},
  doi={10.1016/S0165-0173(98)00056-3}
}

@article{berka2007eeg,
  title={{EEG} correlates of task engagement and mental workload in vigilance, learning, and memory tasks},
  author={Berka, Chris and Levendowski, Daniel J. and Lumicao, Michelle N. and Yau, Alan and Davis, Gene and Zivkovic, Vladimir T. and Olmstead, Richard E. and Tremoulet, Patrice D. and Craven, Patrick L.},
  journal={Aviation, Space, and Environmental Medicine},
  volume={78},
  number={5, Suppl.},
  pages={B231--B244},
  year={2007},
  publisher={Aerospace Medical Association}
}

@article{henrie2015measuring,
  title={Measuring student engagement in technology-mediated learning: A review},
  author={Henrie, Curtis R. and Halverson, Lisa R. and Graham, Charles R.},
  journal={Computers \& Education},
  volume={90},
  pages={36--53},
  year={2015},
  publisher={Elsevier},
  doi={10.1016/j.compedu.2015.09.005}
}

@article{giannakos2019multimodal,
  title={Multimodal data as a means to understand the learning experience},
  author={Giannakos, Michail N. and Sharma, Kshitij and Pappas, Ilias O. and Kostakos, Vassilis and Velloso, Eduardo},
  journal={International Journal of Information Management},
  volume={48},
  pages={108--119},
  year={2019},
  publisher={Elsevier},
  doi={10.1016/j.ijinfomgt.2019.02.003}
}

@article{darnell2019student,
  title={Student engagement, assessed using heart rate, shows no reset following active learning sessions in lectures},
  author={Darnell, Diana K. and Krieg, Paul A.},
  journal={PLoS ONE},
  volume={14},
  number={12},
  pages={e0225709},
  year={2019},
  publisher={Public Library of Science},
  doi={10.1371/journal.pone.0225709}
}

@article{pope1995biocybernetic,
  title={Biocybernetic system evaluates indices of operator engagement in automated task},
  author={Pope, Alan T. and Bogart, Edward H. and Bartolome, Debbie S.},
  journal={Biological Psychology},
  volume={40},
  number={1--2},
  pages={187--195},
  year={1995},
  publisher={Elsevier},
  doi={10.1016/0301-0511(95)05116-3}
}

@article{ko2017sustained,
  title={Sustained attention in real classroom settings: An {EEG} study},
  author={Ko, Li-Wei and Komarov, Oleksii and Hairston, W. David and Jung, Tzyy-Ping and Lin, Chin-Teng},
  journal={Frontiers in Human Neuroscience},
  volume={11},
  pages={388},
  year={2017},
  publisher={Frontiers Media SA},
  doi={10.3389/fnhum.2017.00388}
}

@article{wang2014exploratory,
  title={An exploratory study using inexpensive electroencephalography ({EEG}) to understand flow experience in computer-based instruction},
  author={Wang, Chih-Chien and Hsu, Ming-Chang},
  journal={Information \& Management},
  volume={51},
  number={7},
  pages={912--923},
  year={2014},
  publisher={Elsevier},
  doi={10.1016/j.im.2014.05.010}
}

@article{cavanagh2014frontal,
  title={Frontal theta as a mechanism for cognitive control},
  author={Cavanagh, James F. and Frank, Michael J.},
  journal={Trends in Cognitive Sciences},
  volume={18},
  number={8},
  pages={414--421},
  year={2014},
  publisher={Elsevier},
  doi={10.1016/j.tics.2014.04.012}
}

@article{pfurtscheller1999event,
  title={Event-related {EEG/MEG} synchronization and desynchronization: Basic principles},
  author={Pfurtscheller, Gert and {Lopes da Silva}, Fernando H.},
  journal={Clinical Neurophysiology},
  volume={110},
  number={11},
  pages={1842--1857},
  year={1999},
  publisher={Elsevier},
  doi={10.1016/S1388-2457(99)00141-8}
}

@article{ray1985eeg,
  title={{EEG} alpha activity reflects attentional demands, and beta activity reflects emotional and cognitive processes},
  author={Ray, William J. and Cole, Harry W.},
  journal={Science},
  volume={228},
  number={4700},
  pages={750--752},
  year={1985},
  publisher={American Association for the Advancement of Science},
  doi={10.1126/science.3992243}
}

@article{muhl2014eeg,
  title={{EEG}-based workload estimation across affective contexts},
  author={M{\"u}hl, Christian and Jeunet, Camille and Lotte, Fabien},
  journal={Frontiers in Neuroscience},
  volume={8},
  pages={114},
  year={2014},
  publisher={Frontiers Media SA},
  doi={10.3389/fnins.2014.00114}
}

@article{zhang2021design,
  title={Design and implementation of an {EEG}-based learning-style recognition mechanism},
  author={Zhang, Bingxue and Chai, Chengliang and Yin, Zhong and Shi, Yang},
  journal={Brain Sciences},
  volume={11},
  number={5},
  pages={613},
  year={2021},
  publisher={MDPI},
  doi={10.3390/brainsci11050613}
}

@article{husmann2019another,
  title={Another nail in the coffin for learning styles? {D}isparities among undergraduate anatomy students' study strategies, class performance, and reported {VARK} learning styles},
  author={Husmann, Polly R. and O'Loughlin, Valerie Dean},
  journal={Anatomical Sciences Education},
  volume={12},
  number={1},
  pages={6--19},
  year={2019},
  publisher={Wiley},
  doi={10.1002/ase.1777}
}

@article{knoll2017learning,
  title={Learning style, judgements of learning, and learning of verbal and visual information},
  author={Knoll, Abby R. and Otani, Hajime and Skeel, Reid L. and Van Horn, K. Roger},
  journal={British Journal of Psychology},
  volume={108},
  number={3},
  pages={544--563},
  year={2017},
  publisher={Wiley},
  doi={10.1111/bjop.12214}
}

@article{leite2010attempted,
  title={Attempted validation of the scores of the {VARK}: Learning styles inventory with multitrait--multimethod confirmatory factor analysis models},
  author={Leite, Walter L. and Svinicki, Marilla and Shi, Yuying},
  journal={Educational and Psychological Measurement},
  volume={70},
  number={2},
  pages={323--339},
  year={2010},
  publisher={SAGE Publications},
  doi={10.1177/0013164409344507}
}

@article{dmello2017advanced,
  title={Advanced, analytic, automated ({AAA}) measurement of engagement during learning},
  author={{D'Mello}, Sidney and Dieterle, Ed and Duckworth, Angela},
  journal={Educational Psychologist},
  volume={52},
  number={2},
  pages={104--123},
  year={2017},
  publisher={Taylor \& Francis},
  doi={10.1080/00461520.2017.1281747}
}

@article{dikker2017brain,
  title={Brain-to-brain synchrony tracks real-world dynamic group interactions in the classroom},
  author={Dikker, Suzanne and Wan, Lu and Davidesco, Ido and Kaggen, Lisa and Oostrik, Matthias and McClintock, James and Rowland, Jess and Michalareas, Georgios and Van Bavel, Jay J. and Ding, Mingzhou and Poeppel, David},
  journal={Current Biology},
  volume={27},
  number={9},
  pages={1375--1380},
  year={2017},
  publisher={Elsevier},
  doi={10.1016/j.cub.2017.04.002}
}

@article{gevins1997high,
  title={High-resolution {EEG} mapping of cortical activation related to working memory: Effects of task difficulty, type of processing, and practice},
  author={Gevins, A. and Smith, M. E. and McEvoy, L. and Yu, D.},
  journal={Cerebral Cortex},
  volume={7},
  number={4},
  pages={374--385},
  year={1997},
  publisher={Oxford University Press},
  doi={10.1093/cercor/7.4.374}
}

@article{sawangjai2020consumer,
  title={Consumer grade {EEG} measuring sensors as research tools: A review},
  author={Sawangjai, Phattarapong and Hompoonsup, Supanida and Leelaarporn, Pitshaporn and Kongwudhikunakorn, Supavit and Wilaiprasitporn, Theerawit},
  journal={IEEE Sensors Journal},
  volume={20},
  number={8},
  pages={3996--4024},
  year={2020},
  publisher={IEEE},
  doi={10.1109/JSEN.2019.2962874}
}

@inproceedings{yin2026multimodal,
  title={Multimodal {EEG-IMU} fusion for motor assessment: Leveraging task-dependent complementarity for robustness},
  author={Yin, Zhenan and Pulavarthy, Lalitha Pranathi and Purkayastha, Saptarshi},
  booktitle={2026 IEEE 14th International Conference on Healthcare Informatics (ICHI)},
  pages={204--213},
  year={2026},
  publisher={IEEE},
  doi={10.1109/ICHI69079.2026.00036}
}

@inproceedings{aswar2023generalizability,
  title={Generalizability of human activity recognition machine learning models from non-{Parkinson's} to {Parkinson's} disease patients},
  author={Aswar, Shreyas and Yerrabandi, Vedasree and Moncy, Megha M. and Boda, Sahithi Reddy and Jones, Josette and Purkayastha, Saptarshi},
  booktitle={2023 45th Annual International Conference of the IEEE Engineering in Medicine \& Biology Society (EMBC)},
  pages={1--4},
  year={2023},
  publisher={IEEE},
  doi={10.1109/EMBC40787.2023.10340065}
}

@article{purkayastha2019critical,
  title={Critical components of formative assessment in process-oriented guided inquiry learning for online labs},
  author={Purkayastha, Saptarshi and Surapaneni, Asha K. and Maity, Pallavi and Rajapuri, Anushri S. and Gichoya, Judy W.},
  journal={Electronic Journal of e-Learning},
  volume={17},
  number={2},
  pages={79--92},
  year={2019},
  doi={10.34190/JEL.17.2.02}
}

@inproceedings{sunchu2024lessons,
  title={Lessons learned from {OpenEMR} implementation in graduate health informatics curriculum},
  author={Sunchu, Keerthika and Moncy, Megha M. and Purkayastha, Saptarshi and Fulton, Cathy R.},
  booktitle={AMIA Annual Symposium Proceedings},
  volume={2024},
  pages={1079--1088},
  year={2024},
  organization={American Medical Informatics Association}
}

@article{delorme2004eeglab,
  title={{EEGLAB}: An open source toolbox for analysis of single-trial {EEG} dynamics including independent component analysis},
  author={Delorme, Arnaud and Makeig, Scott},
  journal={Journal of Neuroscience Methods},
  volume={134},
  number={1},
  pages={9--21},
  year={2004},
  publisher={Elsevier},
  doi={10.1016/j.jneumeth.2003.10.009}
}

@article{combrisson2015exceeding,
  title={Exceeding chance level by chance: The caveat of theoretical chance levels in brain signal classification and statistical assessment of decoding accuracy},
  author={Combrisson, Etienne and Jerbi, Karim},
  journal={Journal of Neuroscience Methods},
  volume={250},
  pages={126--136},
  year={2015},
  publisher={Elsevier},
  doi={10.1016/j.jneumeth.2015.01.010}
}

@article{kosmyna2019attentivu,
  title={{AttentivU}: An {EEG}-based closed-loop biofeedback system for real-time monitoring and improvement of engagement for personalized learning},
  author={Kosmyna, Nataliya and Maes, Pattie},
  journal={Sensors},
  volume={19},
  number={23},
  pages={5200},
  year={2019},
  publisher={MDPI},
  doi={10.3390/s19235200}
}
\bibliographystyle{ieeetr}

\end{document}